\documentclass[11pt,a4paper]{emulateapj}

\usepackage{url}
\usepackage[utf8]{inputenc}
\usepackage{fancyref}
\usepackage{hyperref}
\hypersetup{colorlinks,breaklinks,
            urlcolor=[rgb]{0.2,0.2,0.2},  
            linkcolor=[rgb]{0.843,0.098,0.110},  
            citecolor=[rgb]{0,0.4,1.0}}  

\usepackage{natbib}
\newcommand{\teff}{${T}_{\mathrm{eff}}$}
\newcommand{\logg}{$\log{g}$}
\newcommand{\msun}{${M}_{\odot}$}

\newcommand{\muhz}{$\mu$Hz}
\newcommand{\tar}{PG\,1149+057}
\newcommand{\kms}{km\,s$^{-1}$}

\shorttitle{Outbursts in Pulsating White Dwarfs}
\shortauthors{Hermes et al.}

\begin{document}

\title{A SECOND CASE OF OUTBURSTS IN A PULSATING WHITE DWARF OBSERVED BY {\em KEPLER}}
\author{J.~J.~Hermes\altaffilmark{1}, M.~H.~Montgomery\altaffilmark{2}, Keaton~J.~Bell\altaffilmark{2}, P.~Chote\altaffilmark{1}, B.~T.~G\"{a}nsicke\altaffilmark{1}, Steven~D.~Kawaler\altaffilmark{3}, J.~C.~Clemens\altaffilmark{4}, B.~H.~Dunlap\altaffilmark{4}, D.~E.~Winget\altaffilmark{2}, and D.~J.~Armstrong\altaffilmark{1}}

\altaffiltext{1}{Department of Physics, University of Warwick, Coventry\,-\,CV4~7AL, UK}
\altaffiltext{2}{Department of Astronomy, University of Texas at Austin, Austin, TX\,-\,78712, USA}
\altaffiltext{3}{Department of Physics and Astronomy, Iowa State University, Ames, IA\,-\,50011, USA}
\altaffiltext{4}{Department of Physics and Astronomy, University of North Carolina, Chapel Hill, NC\,-\,27599-3255, USA}

\email{j.j.hermes@warwick.ac.uk}

\begin{abstract}

We present observations of a new phenomenon in pulsating white dwarf stars: large-amplitude outbursts at timescales much longer than the pulsation periods. The cool (\teff\,$=11{,}010$\,K), hydrogen-atmosphere pulsating white dwarf \tar\ was observed nearly continuously for more than 78.8\,d by the extended {\em Kepler} mission in {\em K2} Campaign 1. The target showed 10 outburst events, recurring roughly every 8\,d and lasting roughly 15\,hr, with maximum flux excursions up to 45\% in the {\em Kepler} bandpass. We demonstrate that the outbursts affect the pulsations and therefore must come from the white dwarf. Additionally, we argue that these events are not magnetic reconnection flares, and are most likely connected to the stellar pulsations and the relatively deep surface convection zone. \tar\ is now the second cool pulsating white dwarf to show this outburst phenomenon, after the first variable white dwarf observed in the {\em Kepler} mission, KIC~4552982. Both stars have the same effective temperature, within the uncertainties, and are among the coolest known pulsating white dwarfs of typical mass. These outbursts provide fresh observational insight into the red edge of the DAV instability strip and the eventual cessation of pulsations in cool white dwarfs.

\end{abstract}


\section{Introduction}
\label{sec:intro}

The vast majority of all stars in our Galaxy will end up as white dwarfs, which provide important observational boundary conditions on the final stages of stellar evolution. White dwarfs have electron-degenerate cores overlaid by a thin, non-degenerate atmosphere, and passively release their residual thermal energy. When a DA (hydrogen-atmosphere) white dwarf eventually cools below 12,500\,K, it develops a hydrogen partial-ionization zone at its surface. This superficial convection zone throttles heat transport and efficiently drives pulsations \citep{Brickhill91a,GoldreichWu99a}.

The flux changes in variable DA white dwarfs (DAVs, aka ZZ Cetis) are consistent with surface temperature variations caused by non-radial $g$-mode pulsations. By matching the observed pulsation periods to theoretical models, we can reveal considerable detail about the interior of white dwarfs, including their overall mass, temperature, the mass of their outer hydrogen and helium envelopes, their rotation, and even characteristics of their cores (see reviews by \citealt{WinKep08,FontBrass08,Althaus10}).

As white dwarfs cool their convection zones deepen. Since the oscillation modes excited are characterized by the thermal timescale of the driving zone, a deeper convection zone excites long-period pulsations, up to an observed maximum of about 1200\,s for a 0.6\,\msun\ white dwarf \citep{Mukadam06}. Empirical studies of the coolest members of the DAV instability strip are complicated by the relatively long-period and less coherent pulsations \citep{Kanaan02}. However, the {\em Kepler} space telescope provides an unprecedented tool for long-term monitoring of the coolest pulsating white dwarfs.

\begin{figure*}[t]
\centering{\includegraphics[width=0.935\textwidth]{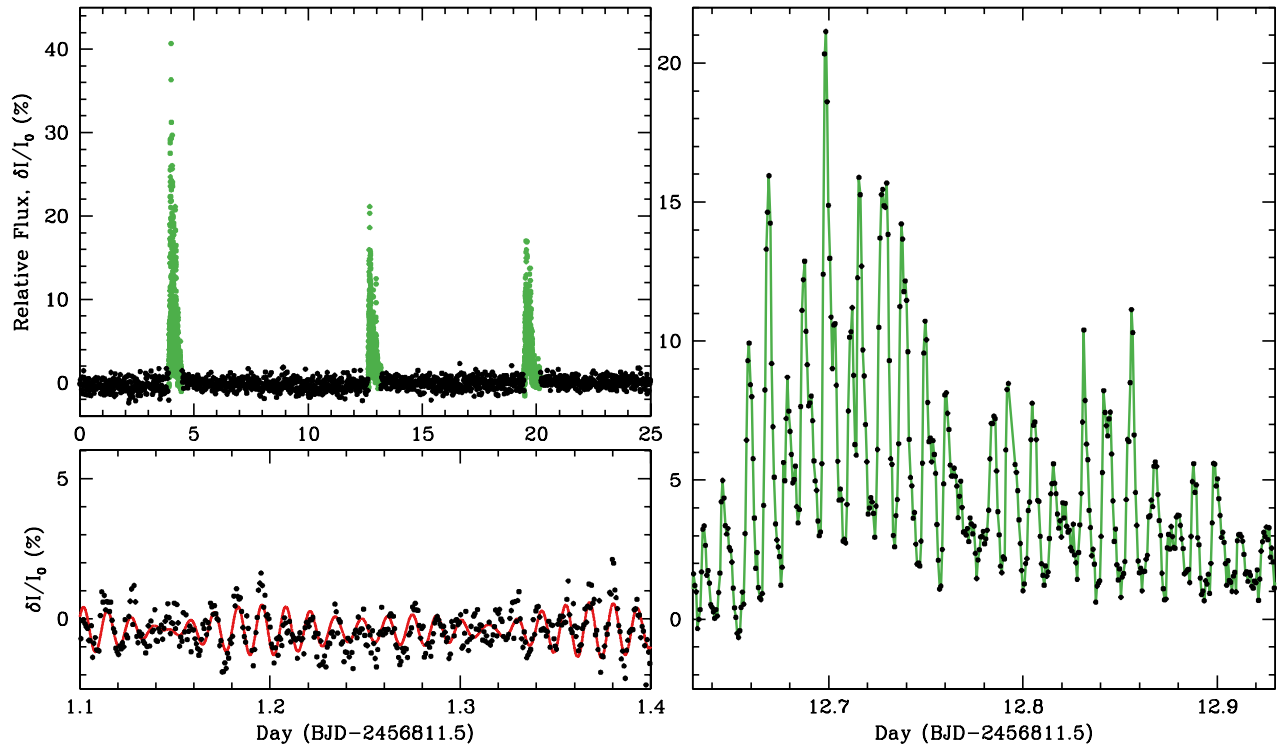}}
\caption{Representative portions of the {\em K2} Campaign 1 light curve of the pulsating white dwarf \tar. The top left panel shows the first 25\,d of observations; three outburst events are denoted in green. The bottom left panel shows 7.2\,hr of data on the second day of {\em K2} observations; the white dwarf pulsations are clearly visible, and underplotted is a best-fit to the three highest-amplitude signals (with periods of 1145.7\,s, 998.1\,s, and 1052.8\,s). The right panel shows 7.2\,hr during the second outburst, with points connected in green. \label{fig:lc}}
\end{figure*}

Ten pulsating white dwarfs were eventually discovered late in the original {\em Kepler} mission, but only two were observed for more than a month before the failure of the second reaction wheel in 2013~May \citep{Hermes11,Greiss14}. Curiously, the first and longest-studied DAV observed by {\em Kepler}, KIC~4552982, showed a completely unprecedented behavior: large-amplitude flux excursions which raised the overall brightness by up to 17\%, recurred on average every 2.7\,d and lasted $4-25$\,hr in duration \citep{Bell15}. These outbursts recurred on a much longer timescale than the pulsations.

The extended {\em Kepler} mission ({\em K2}) affords an exceptional opportunity to extend space-based monitoring for up to 85\,d to a plethora of new (and brighter) white dwarfs \citep{Howell14}. Already {\em K2} has illuminated frequency variability in the cool DAV GD~1212 \citep{Hermes14} and a hot DAV descended from common-envelope evolution \citep{Hermes15}.

The pulsating white dwarf \tar\ was observed for more than 78.8\,d in {\em K2} Campaign 1, and we present here an analysis of the second case of sporadic, high-amplitude outbursts in a cool DAV. In this Letter we present our {\em K2} observations, establish that the outbursts are happening on the white dwarf, and analyze the outbursts and their relation to the stellar pulsations.



\section{{\em K2} Campaign 1 Time-Series Photometry}
\label{sec:k2data}

There is little high-speed photometry on \tar, despite the fact that there are just 15 brighter known DAVs. Pulsations were announced using one very short ($<0.7$\,hr) run in 2005~May by \citet{Voss06}, who present the only previously published light curve.

Our {\em K2} Campaign 1 coverage of \tar\ (EPIC 201806008, $K_p=15.0$\,mag) was requested in short-cadence (exposures every 58.8\,s) and spans from 2014~June~3 00:00:42 UT to 2014~August~20 20:16:36 UT. Every 6\,hr, the {\em Kepler} spacecraft checks its pointing and, if it needs correction, fires its thrusters. This can cause significant discontinuities in the light curve. We have extracted and detrended the photometry using the tools outlined in \citet{Armstrong15}, which compensates for artifacts in the light curve caused by attitude corrections.

Our final data set has $107{,}443$ points, yielding a duty cycle $>$92.8\%. Figure~\ref{fig:lc} shows the first 25\,d of {\em K2} coverage, including the first three of a total of 10 outbursts. Inspecting the light curve in more detail reveals the $1-3$\% peak-to-peak variability caused by the white dwarf pulsations.

\begin{figure*}[t]
\centering{\includegraphics[width=0.935\textwidth]{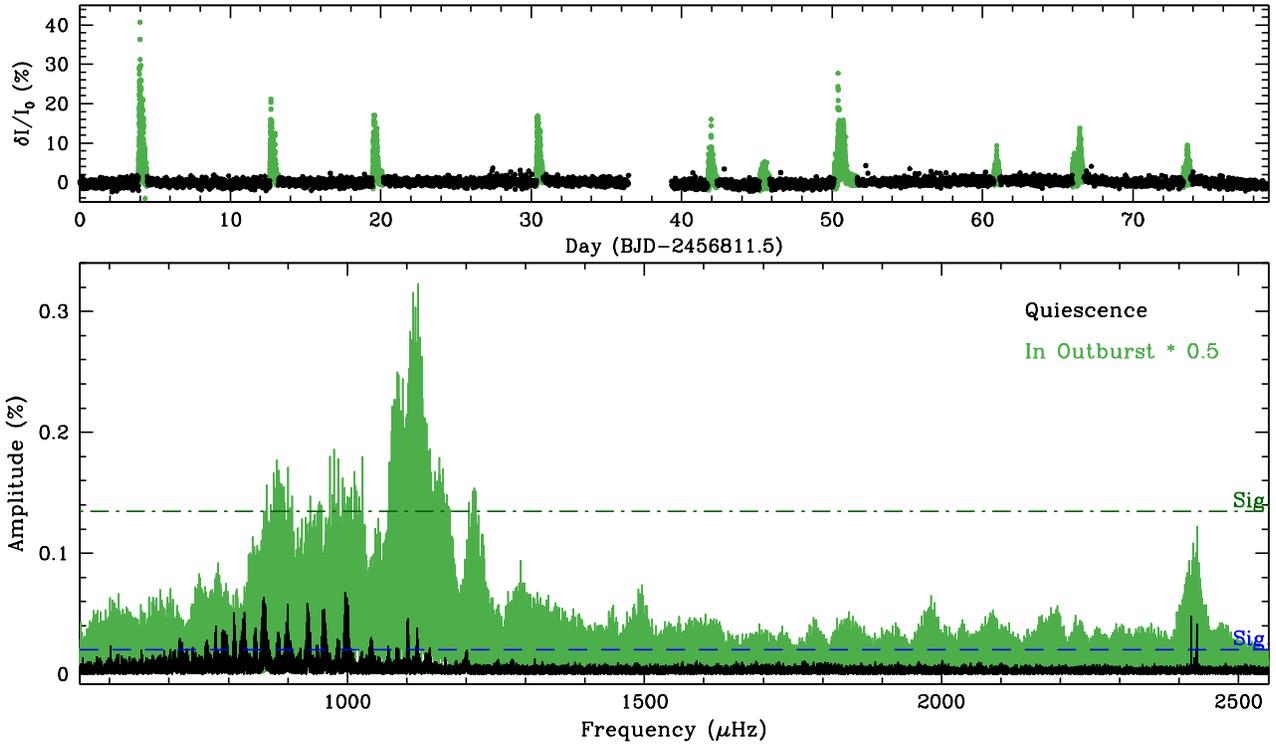}}
\caption{The top panel shows the entire {\em K2} light curve of \tar. We mark the 10 outburst events in green and show the data in quiescence as black points. The bottom panel shows FTs both in (green) and out of outburst (black). Both subsets have pulsations excited above the significance threshold, described in the text and shown as dark green dashed-dotted and blue dashed lines, respectively. Pulsations persist during outburst but have higher amplitudes and shorter periods than in quiescence. \label{fig:ft}}
\end{figure*}

We display the entire {\em K2} light curve in the top panel of Figure~\ref{fig:ft}, marking each outburst in green. The multi-day gap beginning near 2456849\,BJD was caused by a major pointing shift in the spacecraft, which subsequently added minor, long-timescale variations about the mean flux level.

We have inspected movies of the {\em Kepler} images around each of these outbursts using the {\sc k2flix} package \citep{Barentsen15} and see no evidence for passing solar system bodies causing brightness increases. {\em Kepler} has a large plate scale (4 arcsec pixel$^{-1}$), but the nearest source with $r<19.5$\,mag is SDSS\,J115155.09+052959.3, which is more than 1.3\,arcmin to the north, well outside the extracted target pixels.



\section{Archival Spectroscopy and Photometry}
\label{sec:atmospheric}

We describe here the time-averaged atmospheric parameters for \tar, and exclude the possibility that this white dwarf has a flaring, low-mass companion.

To date, the most consistent picture of the parameters of white dwarfs in and around the DAV instability strip has been undertaken by \citet{Gianninas11}. Their one-dimensional, ML2/$\alpha = 0.8$ atmospheric parameters place \tar\ near the empirical red edge of the DAV instability strip, with \teff\ $=11{,}360\pm170$\,K and \logg\ $= 8.21\pm0.05$. \citet{Tremblay13} introduced terms to correct for the three-dimensional (3D) dependence of convection, which refine the final determinations for \tar\ to \teff\ $=11{,}060\pm170$\,K and \logg\ $= 8.06\pm0.05$. This corresponds to a mass of $0.64\pm0.03$\,\msun\ and a distance of $39.3\pm2.4$\,pc using the mass-radius relationship of \citet{Renedo10}.

High-resolution spectra taken as part of the ESO Supernovae Type Ia Progenitor Survey show no evidence of metals from ongoing accretion \citep{Voss06}, and the radial velocities of this white dwarf do not vary to a limit of 4.0\,\kms\ over 4\,hr \citep{MaxtedMarsh99}. We have inspected the Wide Angle Search for Planets survey light curve of \tar, which covers more than 3\,yr but is at the faint end for the survey. Despite binning the data into 30-min intervals, the 1$\sigma$ standard deviation of all points is 0.17\,mag, so we cannot significantly constrain whether outbursts have been observed from previous ground-based studies.

\begin{figure}[b]
\centering{\includegraphics[width=0.975\columnwidth]{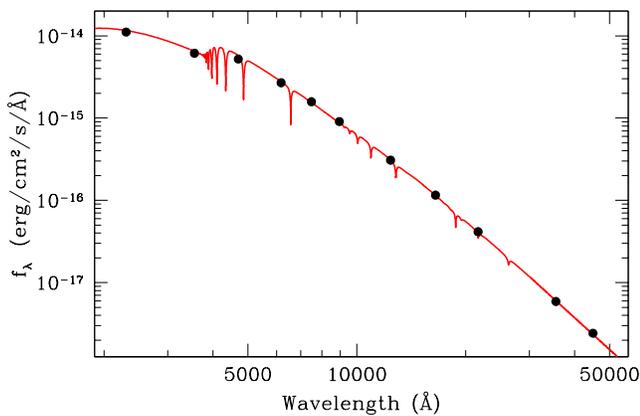}}
\caption{The spectral energy distribution from archival data demonstrates that there is no companion to \tar\ responsible for the outbursts. The data are consistent with an isolated white dwarf (this an $11{,}000$\,K model) from the near-ultraviolet out to the mid-infrared (see text for sources). The 3$\sigma$ photometric uncertainties are smaller than each point. \label{fig:sed}}
\end{figure}

We have examined the wealth of archival photometry on \tar\ and show in Figure~\ref{fig:sed} that the spectral energy distribution (SED) is consistent with a single, isolated white dwarf from the near-ultraviolet out to the mid-infrared. The SED uses photometry from GALEX-NUV \citep{Morrissey07}, SDSS-$ugriz$ and UKIDSS-$JHK$ \citep{Girven11}, and {\em Spitzer} 3.6- and 4.5\,$\mu$m photometry  \citep{Barber12}. Underplotted in Figure~\ref{fig:sed} is an $11{,}000$\,K, \logg\,$=8.0$ DA white dwarf model atmosphere computed by \citet{Koester10}, normalized to the SDSS-$r$ band.

The lack of infrared excess as compared to our white dwarf model rules out any possible companion earlier than L4 using the data of \citet{Cushing05}. Our {\em K2} light curve is therefore composed entirely of \tar, and the outbursts observed are the result of brightening events on the surface of the white dwarf.



\section{Discussion of the Outbursts}
\label{sec:outbursts}

\subsection{Outburst Characteristics}
\label{sec:char}

\begin{figure*}[t]
\centering{\includegraphics[width=0.935\textwidth]{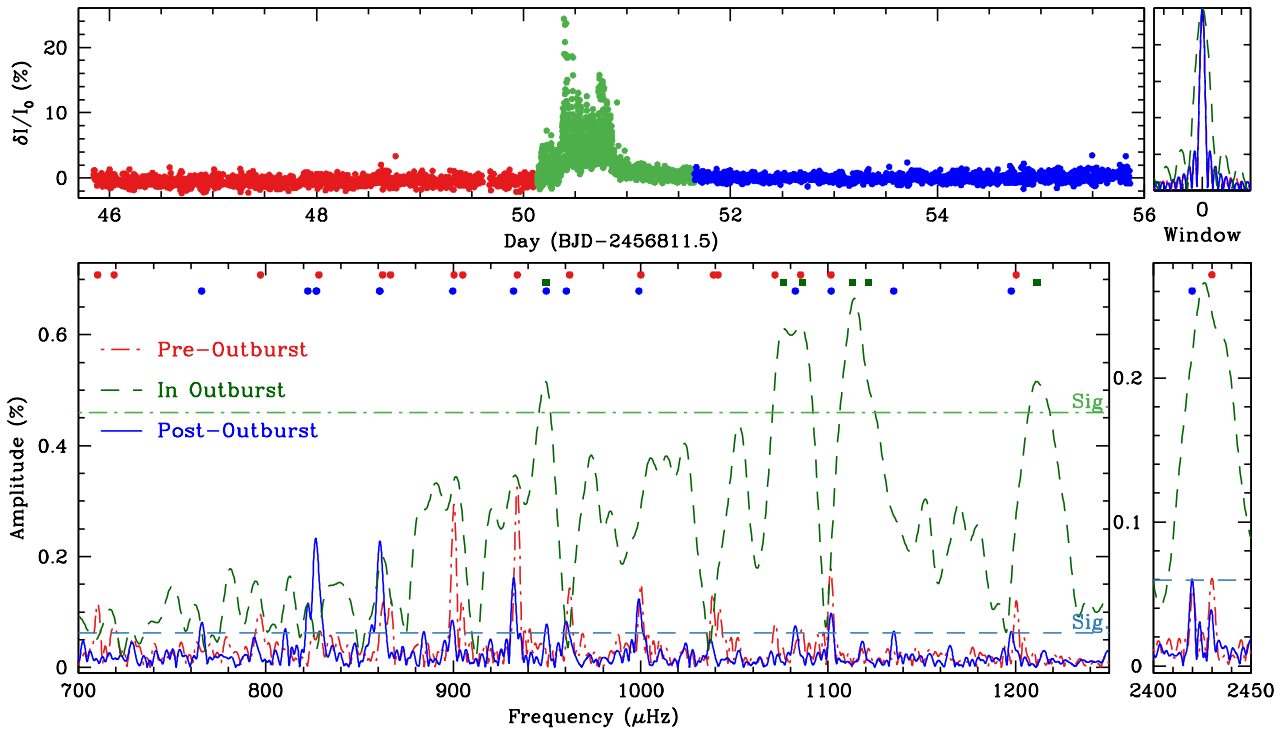}}
\caption{Diagnostics of the pulsations before (red), during (green) and after (blue) the seventh outburst observed on \tar. The top panel shows the {\em K2} light curve, while the bottom panels show the FT of each subset. Signals exceeding the significance thresholds (see text) are marked with appropriate-colored dots and show differences in the excited pulsation spectra. Amplitudes are essentially all lower after outburst, but subsequently grow (see Figure~\ref{fig:rft}). Significant pulsations are also visible during the brightening event (green squares), and appear at shorter periods (higher frequencies) and higher amplitudes. The top right panel shows the window function of each subset. \label{fig:prepost}}
\end{figure*}

The outburst events shown in the top panel of Figure~\ref{fig:ft} confirm a new phenomenon observed in pulsating white dwarfs. Only one other pulsating white dwarf displays similar recurring behavior: The cool DAV KIC~4552982 (WD~J1916+3938), observed for nearly 2.5\,yr in the original {\em Kepler} mission \citep{Bell15}.

That DAV has very similar 3D atmospheric parameters to \tar --- \teff\ $=10{,}860\pm120$\,K and \logg\ $= 8.16\pm0.06$ --- and it also has a very similar pulsation spectrum. The only other reminiscent event seen in a pulsating white dwarf occurred during monitoring of the cool variable DB (helium-atmosphere) white dwarf GD~358, which in 1996 underwent a dramatic change in excited frequencies accompanied by a rapid increase in fractional amplitude, its {\em sforzando} \citep{Montgomery10}.

The outbursts in KIC~4552982 increase its mean flux by between $2-17$\%, last $4-25$\,hr, and recur on an average every 2.7\,d \citep{Bell15}. Everything about the outbursts in \tar\ is scaled up by roughly a factor of three. The outbursts here increase the quiescent flux by between $5-45$\%, have durations on average of 14.5\,hr (ranging from $9.3-36.4$\,hr), and recur on a timescale between $3.5-11.5$\,d, on average every 8\,d. These rough timescales were determined by eye, noting when the point-to-point scatter rose and fell to the quiescent values. With so few events, an autocorrelation function reveals only that most events last more than 12\,hr. There does not appear to be any periodicity to the outbursts.

\subsection{Outbursts Affect the Pulsations}
\label{sec:outburstpulsations}

We see convincing evidence that the pulsations persist on the white dwarf during the outburst, and actually change properties relative to those excited in quiescence. It is important to note that if these outbursts came from a contaminating source the pulsation amplitudes would decrease as a result of flux dilution. That the pulsation amplitudes actually increase proves that the outbursts are on the white dwarf.

The bottom panel of Figure~\ref{fig:ft} shows Fourier transforms (FTs) of the entire light curve, differentiated by data in and out of outburst. Both datasets have peaks in the FT above the respective 3$\sigma$ significance threshold, which was determined by randomly shuffling the fluxes for each point in the light curve and computing the highest peak in the entire resultant FT for $10{,}000$ random permutations (see \citealt{Hermes15}).

Figure~\ref{fig:ft} shows that the pulsations in outburst have shorter periods and higher amplitudes than those in quiescence; the unweighted mean period before, during, and after the seventh outburst goes from 1083.3\,s to 919.7\,s to 1072.7\,s, respectively. If the flux increase during the outbursts are due to an increase in the average effective temperature of the star, the formalism of \citet{Wu99} predicts that the surface amplitudes of the modes should increase, with the enhancement being greater for shorter-period modes. The 1000\,s modes are expected to have growth times of order days \citep{GoldreichWu99a}.


Smoothing the light curve by 3000\,s, over several pulsation cycles, the outbursts increase the mean flux by up to 14\%, which would require a global 750\,K temperature increase from the quiescent effective temperature of $11{,}060$\,K. The pulsations induce even more dramatic, localized temperature excursions.

\begin{figure*}[t]
\centering{\includegraphics[angle=-90,width=0.975\textwidth]{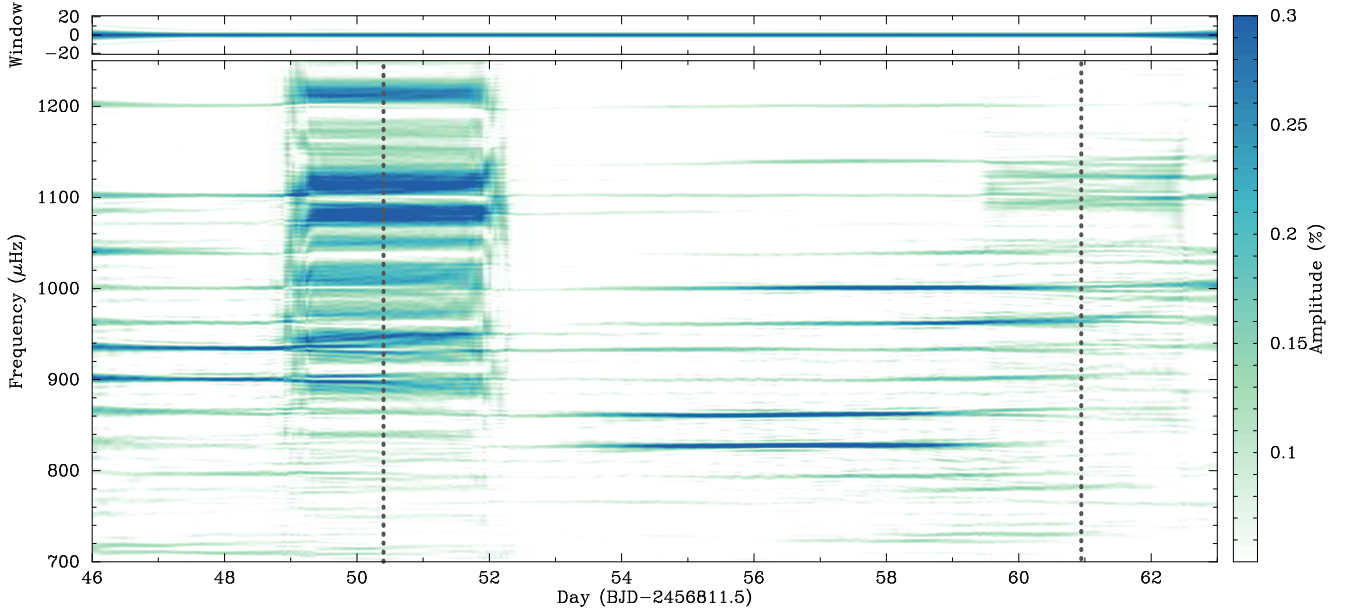}}
\caption{Running FT of the {\em K2} light curve near the seventh outburst shown in Figure~\ref{fig:prepost}. Vertical dotted lines mark the maximum brightness during outburst; our 3-day sliding window smears the events. This plot tracks the post-outburst frequencies, many of which have growing amplitudes until the onset of the eighth outburst peaking near Day\,61. \label{fig:rft}}
\end{figure*}

Figure~\ref{fig:prepost} explores the light curve surrounding and during the seventh observed outburst, between Days $46-56$ from the start of {\em K2} monitoring, using an equal amount of data before and after the outburst. An FT of each subset clearly shows that pulsations are excited to significant amplitudes, even during the outburst.

Many of the significant signals in outburst are not present in quiescence, especially the mode at $949.3\pm0.8$\,\muhz\ ($1053.4\pm0.8$\,s), which decreases in amplitude the longer we monitor it after outburst. Many modes appear to eventually return at roughly the same period after the outburst. However, the amplitude evolution of these pulsations holds key insight.

In essentially all cases in Figure~\ref{fig:prepost}, the amplitudes shown in the subset immediately after the outburst are lower than the pre-outburst subset. There are more data in this post-outburst subset, and we can clearly see many modes growing after the outburst. We explore this amplitude evolution using a running FT, shown in Figure~\ref{fig:rft}. We have computed this diagram by taking an FT across a sliding 3-day window, stepped every hour with 0.05\,\muhz\ frequency resolution.

This time-dependent FT demonstrates that the outburst has depleted considerable pulsation energy. The apparent resetting of amplitudes offers the unprecedented opportunity to actually measure the linear growth rates of pulsation modes in a white dwarf, although that is outside the scope of this Letter.

Finally, the bottom right panel of Figure~\ref{fig:prepost} shows the relatively stable group of three symmetric signals centered at 2425.0\,\muhz\ (the $m=\pm1$ components are visible in this short subset but not the central component). This pulsation triplet arises from rotation of the white dwarf \citep{Pesnell85}, from which we measure an $\ell=1$ frequency splitting of roughly 4.7\,\muhz, which corresponds to a rotation period of roughly 1.2\,d in the asymptotic limit. We reserve a full period determination and detailed asteroseismic analysis of the many pulsations in \tar\ for a future publication.

\subsection{Consideration of Possible Outburst Mechanisms}
\label{sec:mechanisms}

We consider various timescales in white dwarfs to explore different underlying mechanisms for the outbursts. 

White dwarfs have very short dynamical timescales, roughly a few seconds. While this is a global estimate, we expect all processes which are not in hydrostatic equilibrium to locally occur on this or an even shorter timescale. The timescales for the duration and recurrence of the outbursts are clearly much longer than this. This likely rules out magnetic reconnection events as responsible for the outbursts, since the rise times of magnetic flares occur at roughly the dynamical timescale, and the rise times of the outbursts here are of order hours.

On the other hand, thermal processes can occur on longer timescales. To explore this more quantitatively, we have computed a theoretical white dwarf envelope appropriate to this star (11,000\,K, \logg\ = 8.0, ML2/$\alpha$ = 1.0) using the Warsaw -- New Jersey stellar envelope code \citep{Pamyatnykh99}. We find that the thermal timescale corresponding to the average duration of the outbursts, 14.5\,hr, occurs at a depth of $10^{-11.5}$ of the star by mass, placing it slightly deeper but near the base of the convection zone.

We note that the recurrence timescale of roughly 8\,d is mapped to a deeper layer, the outer $10^{-10}$ of the star by mass. It is possible that processes at these depths play a role in the outbursts. However, the recurrence timescale probably rules out runaway thermonuclear events as the source of the temporary brightness increases, since the equilibrium temperature in this region is below 10$^6$\,K, significantly too low for nuclear burning.

An occasional deposition of asteroids onto the surface of the white dwarf could contribute the energy required to match the brightness increases, on average roughly $10^{34}$\,erg integrated over each event, and we now have compelling evidence that disrupted planetesimals frequently pollute the atmospheres of white dwarfs (e.g., \citealt{Koester14}). However, this pollution causes metal absorption, which we can rule out from spectroscopy \citep{Voss06}.

Since the only two known outbursting white dwarfs have very similar effective temperatures and pulsation spectra, it is logical to suspect a connection between the events and the pulsations and the deep convection zone that drives them, especially since we have demonstrated here that the pulsations change before, during, and after outbursts.

We consider a model where the outbursts represent a temporary, rapid reassignment of kinetic energy away from pulsations. It is possible to transfer energy between pulsation modes through resonant mode coupling \citep{Dziembowski82}, which has previously been invoked to explain the amplitude evolution of $\delta$\,Scuti stars in the {\em Kepler} mission \citep{Breger14}. We note that there is still considerable kinetic energy in the modes which are damped in radiative regions of the star as well as from turbulent dissipation in the convection zone.

A theoretical framework exists for parametric instability via mode coupling of white dwarf pulsations, in which energy is rapidly channelled from an observed, overstable parent mode to damped daughter modes \citep{Wu99}. It is possible that the energy of a mode (or multiple modes) grows linearly until it reaches a critical threshold, after which it enters a nonlinear regime and rapidly transfers its energy to resonant daughter modes, or a cascade of resonant daughter modes. These daughter modes may be quickly damped by turbulence in the convection zone and thus deposit their newfound energy there. However, more detailed energetics calculations are required to determine if there is sufficient energy in one (or even multiple) modes to power the observed luminosity increases during outburst.

\subsection{Conclusions}
\label{sec:conclusions}

We present confirmation of a new phenomenon in pulsating white dwarfs by analyzing the second case of outbursts in a cool DAV observed by the {\em Kepler} spacecraft, \tar. We suggest that these outbursts are connected with the pulsations, perhaps as a result of nonlinear mode coupling.

Still, exciting questions remain. Why have we not previously seen outbursts in extensively studied cool DAVs, such as GD\,154 and G29-38 \citep{Pfeiffer96,Kleinman98}? How do the horizontal surface velocities change from quiescence to outburst (e.g., \citealt{vK00})? If resonant mode coupling of pulsations is responsible, what mechanism triggers the outburst? Are the outbursts responsible for the eventual cessation of pulsations and the empirical red edge of the DAV instability strip?

Theory can reasonably reproduce the onset of pulsations observed at the blue edge of the DAV instability strip \citep{Brickhill91b,Wu99}. However, there is still not a clear picture for the red edge where pulsations shut down, which is observed empirically to occur at roughly 10,500\,K for a canonical 0.6\,\msun\ white dwarf \citep{Gianninas14}. Current non-adiabatic calculations predict a much cooler red edge, below 6000\,K \citep{VanGrootel12}.

\citet{Hansen85} proposed that longer-period modes would not be reflected off the surface, setting a condition for the red edge. Subsequently, \citet{Brickhill91b} and \citet{GoldreichWu99b} suggested that turbulent viscosity in the convection zone may damp the longest-period modes and thus define the red edge. However, neither theory provides a complete picture. Perhaps this newly discovered outburst phenomena plays some role in shutting down pulsations in cool DAVs.

Long-timescale monitoring is necessary to catch and constrain future outbursts from cool pulsating white dwarfs, and it is a testament to the utility of the {\em Kepler} instrument to non-planetary science that these events were first discovered thanks to its unblinking gaze. Fortunately, the extended {\em Kepler} mission continues its tour of new {\em K2} fields along the ecliptic every three months, offering the possibility to survey additional cool DAVs in the coming years. We look forward to future theoretical and observational constraints to this unexpected phenomenon.


\acknowledgments

We acknowledge useful discussions with Tom Marsh, Jim Fuller, and Yanqin Wu, and a thoughtful report from an anonymous referee.
We thank Simon Walker for sharing the WASP light curve of \tar\ and thank Detlev Koester for making his DA atmospheric models freely available. 
J.J.H., P.C. and B.T.G. acknowledge funding from the European Research Council under the European Union's Seventh Framework Programme (FP/2007-2013) / ERC Grant Agreement n. 320964 (WDTracer). K.J.B., M.H.M. and D.E.W. gratefully acknowledge the support of the NSF under grant AST-1312983. M.H.M. acknowledges the support of NASA under grant NNX12AC96G. Funding for the {\em Kepler} mission is provided by the NASA Science Mission Directorate.


\begin{thebibliography}{}

\bibitem[Althaus et 
al.(2010)]{Althaus10} Althaus, L.~G., C{\'o}rsico, A.~H., Isern, J., \& Garc{\'{\i}}a-Berro, E.\ 2010, \aapr, 18, 471 

\bibitem[Armstrong et 
al.(2015)]{Armstrong15} Armstrong, D.~J., Kirk, J., Lam, K.~W.~F., et al.\ 2015, \aap, 579, A19 

\bibitem[Barber et al.(2012)]{Barber12} Barber, S.~D., 
Patterson, A.~J., Kilic, M., et al.\ 2012, \apj, 760, 26 

\bibitem[Barentsen(2015)]{Barentsen15} Barentsen, G.\ 2015, 
Astrophysics Source Code Library, 1503.001 

\bibitem[Bell et al.(2015)]{Bell15} Bell, K.~J., Hermes, 
J.~J., Bischoff-Kim, A., et al.\ 2015, arXiv:1506.07878 


\bibitem[Breger 
\& Montgomery(2014)]{Breger14} Breger, M., \& Montgomery, M.~H.\ 2014, \apj, 783, 89 

\bibitem[Brickhill(1991{\natexlab{a}})]{Brickhill91a} Brickhill, A.~J.\ 1991a, 
\mnras, 251, 673 

\bibitem[Brickhill(1991{\natexlab{b}})]{Brickhill91b} Brickhill, A.~J.\ 1991b, 
\mnras, 252, 334 

\bibitem[Cushing et al.(2005)]{Cushing05} Cushing, M.~C., Rayner, 
J.~T., \& Vacca, W.~D.\ 2005, \apj, 623, 1115 

\bibitem[Dziembowski(1982)]{Dziembowski82} Dziembowski, W.\ 1982, AcA, 32, 147 

\bibitem[Fontaine 
\& Brassard(2008)]{FontBrass08} Fontaine, G., \& Brassard, P.\ 2008, \pasp, 120, 1043 

\bibitem[Gianninas et al.(2011)]{Gianninas11} Gianninas, A., 
Bergeron, P., \& Ruiz, M.~T.\ 2011, \apj, 743, 138 

\bibitem[Gianninas et al.(2014)]{Gianninas14} Gianninas, A., 
Dufour, P., Kilic, M., et al.\ 2014, \apj, 794, 35 

\bibitem[Girven et al.(2011)]{Girven11} Girven, J., 
G{\"a}nsicke, B.~T., Steeghs, D., \& Koester, D.\ 2011, \mnras, 417, 1210 

\bibitem[Goldreich 
\& Wu(1999{\natexlab{a}})]{GoldreichWu99a} Goldreich, P., \& Wu, Y.\ 1999a, \apj, 511, 904 

\bibitem[Goldreich 
\& Wu(1999{\natexlab{b}})]{GoldreichWu99b} Goldreich, P., \& Wu, Y.\ 1999b, \apj, 523, 805 

\bibitem[Greiss et al.(2014)]{Greiss14} Greiss, S., 
G{\"a}nsicke, B.~T., Hermes, J.~J., et al.\ 2014, \mnras, 438, 3086 

\bibitem[Hansen et al.(1985)]{Hansen85} Hansen, C.~J., Winget, 
D.~E., \& Kawaler, S.~D.\ 1985, \apj, 297, 544 

\bibitem[Hermes et al.(2011)]{Hermes11} Hermes, J.~J., Mullally, 
F., {\O}stensen, R.~H., et al.\ 2011, \apjl, 741, L16 

\bibitem[Hermes et al.(2014)]{Hermes14} Hermes, J.~J., 
Charpinet, S., Barclay, T., et al.\ 2014, \apj, 789, 85 

\bibitem[Hermes et al.(2015)]{Hermes15} Hermes, J.~J., 
G{\"a}nsicke, B.~T., Bischoff-Kim, A., et al.\ 2015, \mnras, 451, 1701 

\bibitem[Howell et al.(2014)]{Howell14} Howell, S.~B., Sobeck, 
C., Haas, M., et al.\ 2014, \pasp, 126, 398 

\bibitem[Kanaan et 
al.(2002)]{Kanaan02} Kanaan, A., Kepler, S.~O., \& Winget, D.~E.\ 2002, \aap, 389, 896 

\bibitem[Kleinman et al.(1998)]{Kleinman98} Kleinman, S.~J., 
Nather, R.~E., Winget, D.~E., et al.\ 1998, \apj, 495, 424 

\bibitem[Koester(2010)]{Koester10} Koester, D.\ 2010, In Memorie della Societa Astronomica Italiana, 81, 921 

\bibitem[Koester et 
al.(2014)]{Koester14} Koester, D., G{\"a}nsicke, B.~T., \& Farihi, J.\ 2014, \aap, 566, A34 

\bibitem[Maxted 
\& Marsh(1999)]{MaxtedMarsh99} Maxted, P.~F.~L., \& Marsh, T.~R.\ 1999, \mnras, 307, 122 

\bibitem[Montgomery et al.(2010)]{Montgomery10} Montgomery, M.~H., 
Provencal, J.~L., Kanaan, A., et al.\ 2010, \apj, 716, 84 

\bibitem[Morrissey et al.(2007)]{Morrissey07} Morrissey, P., 
Conrow, T., Barlow, T.~A., et al.\ 2007, \apjs, 173, 682 

\bibitem[Mukadam et al.(2006)]{Mukadam06} Mukadam, A.~S., 
Montgomery, M.~H., Winget, D.~E., Kepler, S.~O., 
\& Clemens, J.~C.\ 2006, \apj, 640, 956 

\bibitem[Pamyatnykh(1999)]{Pamyatnykh99} Pamyatnykh, A.~A.\ 1999, AcA, 49, 119 

\bibitem[Pesnell(1985)]{Pesnell85} Pesnell, W.~D.\ 1985, \apj, 
292, 238 

\bibitem[Pfeiffer et 
al.(1996)]{Pfeiffer96} Pfeiffer, B., Vauclair, G., Dolez, N., et al.\ 1996, \aap, 314, 182 

\bibitem[Renedo et al.(2010)]{Renedo10} Renedo, I., Althaus, 
L.~G., Miller Bertolami, M.~M., et al.\ 2010, \apj, 717, 183 


\bibitem[Tremblay et 
al.(2013)]{Tremblay13} Tremblay, P.-E., Ludwig, H.-G., Steffen, M., \& Freytag, B.\ 2013, \aap, 559, AA104 

\bibitem[Van Grootel et 
al.(2012)]{VanGrootel12} Van Grootel, V., Dupret, M.-A., Fontaine, G., et al.\ 2012, \aap, 539, A87 

\bibitem[van Kerkwijk et al.(2000)]{vK00} van Kerkwijk, 
M.~H., Clemens, J.~C., \& Wu, Y.\ 2000, \mnras, 314, 209 

\bibitem[Voss et 
al.(2006)]{Voss06} Voss, B., Koester, D., {\O}stensen, R., et al.\ 2006, \aap, 450, 1061 

\bibitem[Winget 
\& Kepler(2008)]{WinKep08} Winget, D.~E., \& Kepler, S.~O.\ 2008, \araa, 46, 157 

\bibitem[Wu 
\& Goldreich(1999)]{Wu99} Wu, Y., \& Goldreich, P.\ 1999, \apj, 519, 783 

\bibitem[Wu 
\& Goldreich(2001)]{Wu01} Wu, Y., \& Goldreich, P.\ 2001, \apj, 546, 469 


\end{thebibliography}
\end{document}